\documentclass{aamas2012_cameraReady}
\usepackage[protrusion=true,spacing=true]{microtype}
\usepackage[group-separator={,}]{siunitx}
\usepackage{algorithmic}
\usepackage{algorithm}
\usepackage{amsfonts}
\usepackage{graphicx}
\usepackage{cite} % Ordering and compression of citations
\newcommand{\p}{\\[-0.3cm]}

\usepackage[absolute,overlay]{textpos}
\begin{document}

	%%%%%%%
	% Title page
	%%%%%%%

	% Title and authors
	\title{Comparative Evaluation of Multiagent Learning Algorithms \\ in a Diverse Set of Ad Hoc Team Problems}

	\numberofauthors{2}

	\author{
		\alignauthor
		Stefano V. Albrecht \\
		\affaddr{School of Informatics} \\
		\affaddr{University of Edinburgh} \\
		\affaddr{Edinburgh EH8 9AB, UK} \\
		\email{s.v.albrecht@sms.ed.ac.uk}
		\alignauthor
		Subramanian Ramamoorthy \\
		\affaddr{School of Informatics} \\
		\affaddr{University of Edinburgh} \\
		\affaddr{Edinburgh EH8 9AB, UK} \\
		\email{s.ramamoorthy@ed.ac.uk}
	}

	\maketitle

	\begin{textblock}{55}(5,3)
		\small This arXiv version of the original paper published in AAMAS 2012 uses an expanded title to spell out ``MAL'', and is otherwise identical.
	\end{textblock}

	% Abstract
	\begin{abstract}
This paper is concerned with evaluating different multiagent learning (MAL) algorithms in problems where individual agents may be heterogenous, in the sense of utilising different learning strategies, without the opportunity for prior agreements or information regarding coordination. Such a situation arises in \emph{ad hoc team} problems, a model of many practical multiagent systems applications. Prior work in multiagent learning has often been focussed on homogeneous groups of agents, meaning that all agents were identical and a priori aware of this fact. Also, those algorithms that are specifically designed for ad hoc team problems are typically evaluated in teams of agents with fixed behaviours, as opposed to agents which are adapting their behaviours. In this work, we empirically evaluate five MAL algorithms, representing major approaches to multiagent learning but originally developed with the homogeneous setting in mind, to understand their behaviour in a set of ad hoc team problems. All teams consist of agents which are continuously adapting their behaviours. The algorithms are evaluated with respect to a comprehensive characterisation of repeated matrix games, using performance criteria that include considerations such as attainment of equilibrium, social welfare and fairness. Our main conclusion is that there is no clear winner. However, the comparative evaluation also highlights the relative strengths of different algorithms with respect to the type of performance criteria, e.g., social welfare vs. attainment of equilibrium.\p
	\end{abstract}

	% Category, terms, keywords

	%%%%
	% Text
	%%%%

	% Introduction
	\section{Introduction} \label{sec:introduction}

\noindent Game theory provides a mathematically well defined framework for the analysis of multiagent interactive decision making problems. A game consists of a number of players, a set of actions for each player, and a payoff function for each player. A large portion of the theory is developed under the assumption that each player knows the structure of the game, i.e. the action sets and payoff functions of all players. This is known as a \emph{complete information} game. A more complicated setting arises if the players have only partial information about the structure of the game, e.g. no player knows the payoff function of any other player. This is called an \emph{incomplete information} game. While this type of game has been studied for over half a century (notably by Harsanyi \cite{h1962,h1967,h1968a,h1968b}), the problem of incomplete information in multiagent learning (MAL) has received relatively lesser attention. In a version of this setting, called the \emph{ad hoc team} problem \cite{skkr2010}, one seeks to design an autonomous agent which is able to collaborate efficiently with a previously unknown group of agents, in the absence of any prior coordination between the agent and its counterparts chosen in an ad hoc way.\p

This problem is motivated by numerous practical and important applications. As we increasingly employ autonomous agents in a growing number of areas, ranging from teams of robots in automated factories to internet trading agents, it is likely that these agents will have to collaborate with each other in nontrivial ways. As teams are augmented and continually modified over a long lifetime of operation, we would like agents to be able to learn how to collaborate efficiently with other agents, despite not knowing -- a priori -- who they are. An example of such interaction is given by Stone and Kraus in \cite{sk2010}. Another example comes from the domain of human-robot interaction in which a robot has to collaborate with a human, especially when the robot is not given any specific information about the characteristics of the human participant it must collaborate with.\p

The literature on the ad hoc team problem includes a number of proposals for models of interaction and learning. One approach is to try to learn to categorise other agents according to their behaviour or to use a set of role templates to constrain the interactions, e.g. \cite{bsk2011,gas2011}. While these are useful ideas, it is worth noting that many realistic interactive decision making problems involve one further feature not captured in these models -- multiple agents that are simultaneously trying to learn and adapt their behaviour. In such a setting, the environment is inherently nonstationary \cite{y2004} and the interactive behaviour needs more careful treatment. Motivated by this general issue, this paper reports on an empirical study into the behaviour of multiagent learning algorithms in ad hoc teams.\p

There are two open issues with current MAL algorithms in the literature. First, we note that most MAL algorithms (e.g. \cite{cb1998,uv1997,bs2007,bv2002,hm2001,hw2003,l2001,cs2003,gh2003}) were primarily evaluated in homogeneous groups of agents, meaning that all agents in the group were identical, and all agents were a priori aware of this fact. Second, virtually all MAL algorithms which were specifically designed for ad hoc team problems (e.g. \cite{sk2010,skr2010,gas2011,bsk2011,wzc2011}) were primarily evaluated in teams of agents with fixed behaviours. In this work, we evaluate five MAL algorithms of the first sort in a series of ad hoc team problems. In addition, every team consists of agents which are continuously adapting their behaviours. The algorithms are evaluated in a comprehensive set of repeated games, ranging from games in which the players agree on what is most preferred, to games in which the players disagree on what is most preferred. Our performance criteria include the convergence rate, the final expected payoff, social welfare and fairness, and the rates of several solution types.\p

Our intention was to identify those approaches covered by our selection of algorithms which may be better suited for ad hoc team problems. However, as we will show, our results indicate that there is no clear favourite among the algorithms.\p

The remainder of this paper is structured as follows: Section~\ref{sec:exp-setup} describes the experimental setup, including algorithms, games, and performance criteria. Section~\ref{sec:exp-results} presents and analyses the results of our experiments. Section~\ref{sec:rel-work} discusses related work and Section~\ref{sec:conclusion} concludes our work.\p

	% Experimental setup
	\section{Experimental setup} \label{sec:exp-setup}

		% Algorithms
		\subsection{Algorithms} \label{sec:algorithms}

\noindent Our selection of algorithms is motivated by the range of approaches it covers. We tested two algorithms that model their opponents and three that do not model their opponents:\p

\begin{itemize}
	\item Joint Action Learning (JAL) \cite{cb1998,uv1997}
	\item Conditional Joint Action Learning (CJAL) \cite{bs2007}
	\item Win or Learn Fast with PHC (WOLF-PHC) \cite{bv2002}
	\item Modified Regret-Matching\footnote{We use the \emph{Hannan-consistent} version of RegMat. This means that for each action $\hat{a}_i$, the Hannan regret $R_t(\hat{a}_i) = \frac{1}{t} \sum_{\tau = 1}^t u_i(\hat{a}_i,a_{-i}^{\tau}) - \frac{1}{t} \sum_{\tau = 1}^t u_i(a^{\tau})$ will be $\leq 0$ as $t \rightarrow \infty$, where $a^{\tau}$ denotes the joint action played at time $\tau$ (see \cite{hm2001}).} (RegMat) \cite{hm2001}
	\item Nash Q-Learning (NashQ) \cite{hw2003}
\end{itemize}

JAL tries to model its opponents by learning their marginal action probabilities. It uses these probabilities to compute the expected payoffs of all of its actions. CJAL extends JAL in that it learns the action probabilities of its opponents conditioned on its own actions. WOLF-PHC uses a hill climbing method in the space of mixed strategies to find an optimal strategy. RegMat minimises the regret it feels for not having played any other actions. Finally, NashQ tries to learn the payoff distributions of all agents and plays a Nash equilibrium strategy in each state, regardless of the actual behaviour of its opponents.\p

		% Games
		\subsection{Games} \label{sec:games}

\noindent The algorithms are evaluated in a range of \emph{repeated games}. A repeated game $\Gamma$ is a tuple $(N,(A_i)_{i \in N},(u_i)_{i \in N})$, where $N = \left\{ 1,...,n \right\}$ is the set of players (or agents), $A_i$ is the set of actions available to player $i$, and $u_i : A \rightarrow \mathbb{R}$ is the payoff function of player $i$, where $A = A_1 \times ... \times A_n$. In each repetition of the game, each player $i \in N$ simultaneously chooses an action $a_i \in A_i$ and receives the payoff $u_i(a_1,...,a_n)$. Each player $i$ chooses its actions based on a strategy $\pi_i : A_i \rightarrow [0,1]$, which is a probability distribution over the set $A_i$. A strategy $\pi_i$ is called a \emph{pure} strategy if $\pi_i(a_i) = 1$ for some $a_i \in A_i$. A strategy is called a \emph{mixed} strategy if it is not a pure strategy. Given a strategy profile $\pi = (\pi_1,...,\pi_n)$, the expected payoff to player $i$ is defined as $U_i(\pi) = \sum_{a_1,...,a_n} \pi_1(a_1) * ... * \pi_n(a_n) * u_i(a_1,...,a_n)$.\p

Our experiments are divided into three parts. The first two parts evaluate the algorithms in the set of all structurally distinct strictly ordinal $2 \times 2$ no-conflict and conflict games, respectively (based on \cite{rg1966}). A $m_1 \times ... \times m_n$ game is one in which there are $n$ players, each of which with $m_i$ actions. In an \emph{ordinal} game, each player ranks each of the $k = \prod_i m_i$ possible outcomes from 1 (least preferred) to $k$ (most preferred). An ordinal game is called \emph{strictly ordinal} if no two outcomes have the same rank. An ordinal game is called a \emph{no-conflict} game if all players have the same set of most preferred outcomes, otherwise it is called a \emph{conflict} game. Finally, a set of games is said to be structurally distinct of no game in the set can be reproduced by any transformation of any other game in the set. Possible transformations include interchanging the rows, columns, players, and any combination of these in the payoff matrix of the game.\p

We divided the games into no-conflict and conflict games because these define two distinct levels of difficulty. In a no-conflict game, it is relatively easy to arrive at a solution that is best for all players since all players have the same most preferred outcomes. However, in a conflict game, there is no such outcome. Therefore, the agents will have to arrange some form of a compromise. This requires reliable coordination mechanisms, especially in the context of ad hoc teams.\p

The third part of our experiments uses a modified version of the evaluation procedure proposed by Stone et al.~\cite{skkr2010}.~The procedure tests the algorithms in a number of randomly generated strictly ordinal $2 \times 2 \times 2$ games. Each team may contain multiple agents of the same type. Every game is repeated for each algorithm under the same conditions. Algorithm~\ref{alg:eval} shows the pseudo-code of the procedure.\p

\begin{algorithm}[t]
	\caption{Modified evaluation procedure}
	\label{alg:eval}
	\begin{algorithmic}
		\STATE \textbf{Initialise:} Empty vector $M_i$ for each agent $i \in N$
		\LOOP
			\STATE Randomly generate a strictly ordinal $2 \times 2 \times 2$ game $\Gamma$
			\STATE Randomly generate a team $B$ from $N$ with $|B| = 2$
			\FORALL{$i \in N$}
				\STATE Play $\Gamma$ with agents $\left\{ i \right\} \cup B$, where agent $i$ is player 1
				\STATE Compute metrics for agent $i$ and store them in $M_i$
			\ENDFOR
		\ENDLOOP
		\STATE Return averaged metrics $avg(M_i)$ for each $i \in N$
	\end{algorithmic}
\end{algorithm}

		% Performance criteria
		\subsection{Performance criteria}

\noindent This section provides definitions of the performance criteria used in our experiments. The definitions are based on the notion of \emph{plays} of repeated games. A play $P_{\Gamma}$ of a repeated game $\Gamma$ is a tuple $((\pi^t)_{t = 1,...,t_f},(a^t)_{t = 1,...,t_f},(r^t)_{t = 1,...,t_f})$, where $t \in \left\{ 1,...,t_f \right\}$ denotes the time, $t_f$ denotes the final time of the play, $\pi^t = (\pi_1^t,...,\pi_n^t)$ denotes the strategy profile at time $t$, $a^t = (a_1^t,...,a_n^t)$ denotes the joint action at time $t$, and $r^t = (r_1^t,...,r_n^t)$ denotes the joint payoff at time $t$.\p

			% Convergence rate
			\subsubsection{Convergence rate}

\noindent The convergence rate of an agent is defined as the percentage of plays in which the agent converged. Let $P_{\Gamma}$ be a play of some repeated game $\Gamma$. We say that agent $i$ converged in $P_{\Gamma}$ if its strategies $\pi_i^t$ stay within a tolerance bound of $\pm 5\%$ in the final 20\% of the play. Formally, agent $i$ converged in play $P_{\Gamma}$ if for all $t \in \left\{ t_s,...,t_f \right\}$ and $j \in \left\{ 1,...,m_i \right\}$, where $t_s = 0.8 \, t_f$, we have $|\pi_i^{t_s}(j) - \pi_i^t(j)| \leq 0.05$.\p

A high convergence rate is not necessarily better than a low convergence rate. However, we argue that it may be useful if an agent has a high convergence rate as this means that the other agents will have to adapt to one less opponent (that is, once the agent has converged).\p

			% Final expected payoff
			\subsubsection{Final expected payoff}

\noindent The final expected payoff of an agent is an approximation of the agent's expected payoff after having learned for $t_f$ repetitions. The approximation is based on the final 20\% of the play. Let $P_{\Gamma}$ be a play of some repeated game $\Gamma$. The final expected payoff of agent $i$ in play $P_{\Gamma}$ is formally defined as $\bar{r}_i = \frac{1}{t_f - t_s + 1} \sum_{t = t_s}^{t_f} r_i^t$, where $t_s = 0.8 \, t_f$. This is an important metric because it is a major indicator of the algorithm's individual performance.\p

			% Social welfare and fairness
			\subsubsection{Social welfare and fairness}

\noindent Let $P_{\Gamma}$ be a play of some repeated game $\Gamma$. We define the social welfare and fairness of play $P_{\Gamma}$, respectively, as the sum and product of the final expected payoffs $\bar{r}_i$ of all agents $i \in N$. Formally, this corresponds to $\sum_{i = 1}^n \bar{r}_i$ for the social welfare, and $\prod_{i = 1}^n \bar{r}_i$ for the social fairness. These metrics are useful for the assessment of an algorithm's team performance.\p

			% Rate of different solution types
			\subsubsection{Rate of different solution types}

\noindent We measure the rates of four different solution types. The definitions in the following sections rely on the notion of the \emph{averaged final profile}. Let $P_{\Gamma}$ be a play for some repeated game $\Gamma$. The averaged final profile (or AFP) of $P_{\Gamma}$ is the average of all strategy profiles in the final 20\% of the play. Formally, we denote the AFP by $\bar{\pi} = (\bar{\pi}_1,...,\bar{\pi}_n)$, where $\bar{\pi}_i = \frac{1}{t_f - t_s + 1} \sum_{t = t_s}^{t_f} \pi_i^t$ and $t_s = 0.8 \, t_f$. We use the final 20\% of the play to approximate mixed strategies for agents that play pure strategies only (such as JAL and CJAL).\p

				% Nash equilibrium rate
				\paragraph{Nash equilibrium rate}

\noindent The Nash equilibrium (NE) rate is defined as the percentage of plays in which the AFP constitutes a Nash equilibrium. A strategy profile $\pi = (\pi_1,...,\pi_n)$ is a Nash equilibrium if $U_i(\pi_1,...,\pi_i,...,\pi_n) \geq U_i(\pi_1,...,\hat{\pi}_i,...,\pi_n)$ for all players $i$ and all strategies $\hat{\pi}_i$. Given a play $P_{\Gamma}$ of a repeated game $\Gamma$, we determine if the play resulted in a Nash equilibrium by solving the following linear programme for each player $i \in N$: \\

\hspace{1.3cm} \begin{tabular}{rl}
	Maximise: & $U_i(\bar{\pi}_1,...,\pi_i,...,\bar{\pi}_n)$ \\
	Subject to: & $\forall j \in A_i : \pi_i(j) \geq 0$ \\
	& $\sum_{j \in A_i} \pi_i(j) = 1$
\end{tabular}

\vspace{0.3cm}

We denote the optimised profile for player $i$ by $\pi^i$. If, for any $i \in N$, the expected payoff under the optimised profile, $U_i(\pi^i)$, exceeds the expected payoff under the AFP, $U_i(\bar{\pi})$, by more than 5\%, then we conclude that the play $P_{\Gamma}$ did not result in a Nash equilibrium. Formally, we define $P_{\Gamma}$ to result in a Nash equilibrium if $\forall i \in N : \frac{U_i(\pi^i)}{U_i(\bar{\pi})} \leq 1.05$.\p

				% Pareto optimality rate
				\paragraph{Pareto optimality rate}

\noindent The Pareto optimality (PO) rate is defined as the percentage of plays in which the AFP is Pareto-optimal. A strategy profile $\pi$ is Pareto-optimal if there is no other profile $\hat{\pi}$ such that $\forall i \in N : U_i(\hat{\pi}) \geq U_i(\pi)$ and $\exists i \in N : U_i(\hat{\pi}) > U_i(\pi)$. Similar to~\cite{bs2007}, we determine if a profile is Pareto-optimal by measuring its orthogonal distance to the \emph{Pareto front}.\p

Consider a repeated game $\Gamma$. The space of possible expected (joint) payoffs of $\Gamma$ is a convex polytope in $\mathbb{R}^n$ that has one dimension for each player $i \in N$. It is defined as the convex hull of all joint payoffs $(u_1(a),...,u_n(a))$ for all joint actions $a \in A_1 \times ... \times A_n$. Each point in this payoff polytope corresponds to a tuple of expected payoffs $(U_1(\pi),...,U_n(\pi))$ for some profile $\pi = (\pi_1,...,\pi_n)$. The \emph{Pareto front} of a payoff polytope $\Phi$ is defined as the set of Pareto-optimal faces of $\Phi$. A face $\phi$ of $\Phi$ is Pareto-optimal if the corresponding strategy profiles of all points on $\phi$ are Pareto-optimal. Now, given a play $P_{\Gamma}$ of $\Gamma$, we say that it results in a Pareto-optimal solution if the minimal orthogonal distance of its AFP, $\bar{\pi}$, when projected onto the payoff polytope $\Phi$ of $\Gamma$, to the Pareto front of $\Gamma$ is not greater than 0.1. Tests indicate that this value works well for ordinal games.\p

\begin{figure}[h]
	\centering
	\begin{tabular}{cc}
		\begin{minipage}{0.1\textwidth}
			\begin{flushright}
				\includegraphics[width=2.9\textwidth]{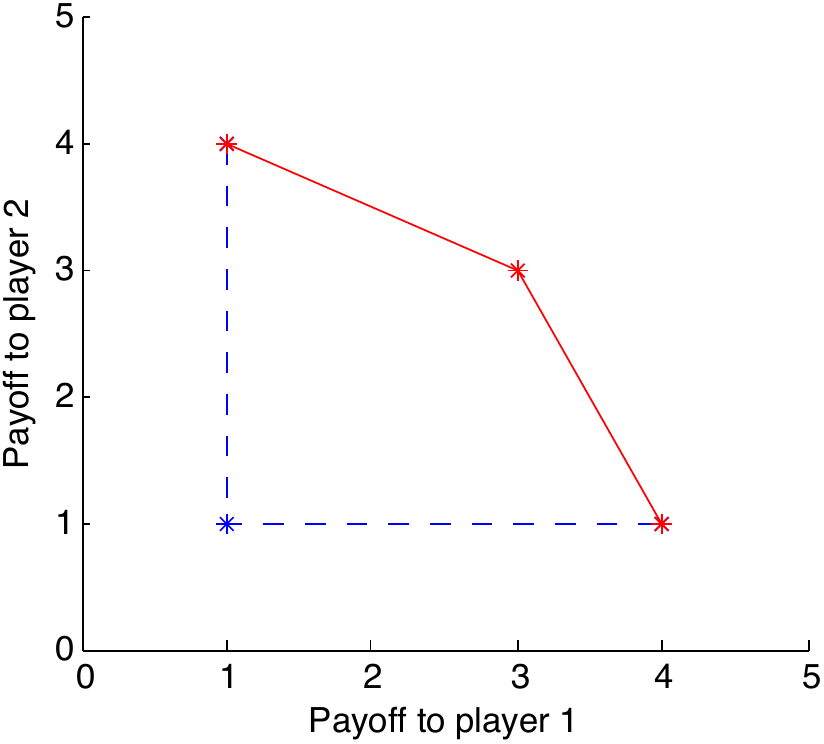}
			\end{flushright}
		\end{minipage}
		&
		\begin{minipage}{0.22\textwidth}
			\begin{flushright}
				\begin{tabular}{|cc|}
					\hline
					1, 1 & 4, 1 \\
					1, 4 & 3, 3 \\
					\hline
				\end{tabular}
			\end{flushright}
		\end{minipage}
	\end{tabular}
	\caption{Example of a Pareto front.}
	\label{fig:pfront-example}
\end{figure}

Figure~\ref{fig:pfront-example} shows the payoff polytope of a $2 \times 2$ game. The payoff table of the game is given in the figure. Player 1 chooses the row and player 2 chooses the column. The elements of the table contain the payoffs to player 1 and 2, respectively.\p

The asterisks in the plot mark the four payoff pairs. The edges correspond to the faces of the polytope. Together, they define the convex hull of all payoff pairs. The solid edges show the Pareto-optimal faces of the polytope. Together, they define the Pareto front of the payoff polytope. This means that every strategy profile whose expected payoff is on\footnote{Or, as in our case, \emph{close} to the Pareto front.} the Pareto front constitutes a Pareto-optimal profile.\p

% +++++ Tables +++++ %
\begin{table*}[t]
	\small
	\centering
	\begin{tabular}{|c|cc|cc|cccc|}
		\hline
		\textbf{Agent} & \textbf{Conv.} & \textbf{Fexp.} & \textbf{Welfare} & \textbf{Fairness} & \textbf{NE} & \textbf{PO} & \textbf{WO} & \textbf{FO} \\
		\hline
		JAL			& 1 & 3.9866 & 7.9720 & 15.9063 & 1 & 0.9920 & 0.9920 & 0.9920 \\
		CJAL		& 1 & 3.9831 & 7.9663 & 15.8874 & 1 & 0.9897 & 0.9897 & 0.9897 \\
		WOLF-PHC	& 0.9996 & 3.9449 & 7.8908 & 15.6426 & 1 & 0.9638 & 0.9638 & 0.9638 \\
		RegMat		& 0.9990 & 3.9107 & 7.8170 & 15.3906 & 0.9954 & 0.9457 & 0.9457 & 0.9457 \\
		NashQ		& 0.9987 & 3.9840 & 7.9733 & 15.9144 & 0.9954 & 0.9939 & 0.9939 & 0.9939 \\
		\hline
	\end{tabular}
	\caption{Results for no-conflict games.}
	\label{tab:res-noconflict}
\end{table*}

\begin{table*}[t]
	\small
	\centering
	\begin{tabular}{|c|cc|cc|cccc|}
		\hline
		\textbf{Agent} & \textbf{Conv.} & \textbf{Fexp.} & \textbf{Welfare} & \textbf{Fairness} & \textbf{NE} & \textbf{PO} & \textbf{WO} & \textbf{FO} \\
		\hline
		JAL			& 0.8901 & 3.0140 & 6.0592 & 8.9997 & 0.8982 & 0.7781 & 0.7021 & 0.6164 \\
		CJAL		& 0.9456 & 3.0326 & 6.0978 & 9.0900 & 0.8470 & 0.8050 & 0.7184 & 0.6250 \\
		WOLF-PHC	& 0.9430 & 3.0392 & 6.0620 & 9.0517 & 0.9047 & 0.7636 & 0.6992 & 0.6142 \\
		RegMat		& 0.8673 & 3.0313 & 6.0368 & 8.9610 & 0.8946 & 0.7662 & 0.7000 & 0.6109 \\
		NashQ		& 0.9990 & 3.0446 & 6.0667 & 9.0755 & 0.8722 & 0.7767 & 0.6946 & 0.6097 \\
		\hline
	\end{tabular}
	\caption{Results for conflict games.}
	\label{tab:res-conflict}
\end{table*}

\begin{table*}[t]
	\small
	\centering
	\begin{tabular}{|c|cc|cc|cccc|}
		\hline
		\textbf{Agent} & \textbf{Conv.} & \textbf{Fexp.} & \textbf{Welfare} & \textbf{Fairness} & \textbf{NE} & \textbf{PO} & \textbf{WO} & \textbf{FO} \\
		\hline
		JAL			& 0.854 & 5.7964 & 17.2174 & 193.1478 & 0.804 & 0.712 & 0.466 & 0.396 \\
		CJAL		& 0.922 & 5.7856 & 17.3521 & 196.1594 & 0.760 & 0.742 & 0.486 & 0.418 \\
		WOLF-PHC	& 0.918 & 5.7400 & 17.1956 & 193.0255 & 0.824 & 0.676 & 0.442 & 0.388 \\
		RegMat		& 0.852 & 5.7290 & 17.2315 & 193.9011 & 0.844 & 0.708 & 0.438 & 0.392 \\
		NashQ		& 0.980 & 5.7562 & 17.2452 & 193.6470 & 0.790 & 0.734 & 0.452 & 0.388 \\
		\hline
	\end{tabular}
	\caption{Results for random games.}
	\label{tab:res-random}
\end{table*}
% +++++ Tables +++++ %

				% Welfare/Fairness optimality rate
				\paragraph{Welfare/Fairness optimality rate}

\noindent The welfare optimality (WO) and fairness optimality (FO) rates are defined as the percentage of plays in which the AFP is welfare-optimal and fairness-optimal, respectively. Given a play $P_{\Gamma}$ of a repeated game $\Gamma$, we say that it results in a welfare-optimal (fairness-optimal) solution if the welfare (fairness) of its AFP is not more than 5\% lower than the maximum welfare (fairness) of $\Gamma$. The maximum welfare (fairness) of a game is the highest possible welfare (fairness) achievable by any strategy profile.\p

Given a game $\Gamma$, we compute its maximum welfare by solving the following non-linear optimisation problem: \\

\hspace{0.75cm} \begin{tabular}{rl}
	Maximise:	 & $\sum_{i \in N} U_i(\pi)$ \\
	Subject to: & $\forall i \in N ~\forall j \in A_i : \pi_i(j) \geq 0$ \\
	& $\forall i \in N : \sum_{j \in A_i} \pi_i(j) = 1$
\end{tabular}

\vspace{0.3cm}

Similarly, we compute its maximum fairness by solving the following non-linear optimisation problem: \\

\hspace{0.75cm} \begin{tabular}{rl}
	Maximise:	 & $\prod_{i \in N} U_i(\pi)$ \\
	Subject to: & $\forall i \in N ~\forall j \in A_i : \pi_i(j) \geq 0$ \\
	& $\forall i \in N : \sum_{j \in A_i} \pi_i(j) = 1$
\end{tabular}

\vspace{0.3cm}

We denote the optimised profile of the first problem by $\pi^w$ and the one of the second problem by $\pi^f$. Then, for a given play $P_{\Gamma}$, we say that it resulted in a welfare-optimal solution if $\frac{W(\pi^w)}{W(\bar{\pi})} \leq 1.05$, where $W(\pi) = \sum_{i \in N} U_i(\pi)$. Similarly, we say that it resulted in a fairness-optimal solution if $\frac{F(\pi^f)}{F(\bar{\pi})} \leq 1.05$, where $F(\pi) = \prod_{i \in N} U_i(\pi)$.\p

		% Parameter settings and selection strategies
		\subsection{Parameter settings and selection strategies}

\noindent With the exception of RegMat, all algorithms are based on Q-learning \cite{wd1992}. This means that they use a table $Q$ to store estimated values for each joint action $a \in A$. The values are updated using a formula of the form $Q(a) \gets (1-\alpha) Q(a) + \alpha \, r$, where $r$ is the payoff to the algorithm, and $\alpha$ is the learning rate. We use a constant learning rate of $\alpha = 0.1$ for all algorithms throughout all experiments. This violates the standard conditions for stochastic approximation \cite{jjs1994}, but enables the algorithms to learn continuously.\p

Furthermore, all algorithms except RegMat use a selection strategy to choose their actions. We use an $\epsilon$-greedy strategy for all algorithms. Therein, the algorithm chooses a random action with probability $\epsilon$, and the greedy action (i.e. the action that is currently believed to have the highest expected payoff) with probability $1 - \epsilon$. We use a constant exploration rate of $\epsilon = 0.05$ for all algorithms. RegMat chooses its actions similar to $\epsilon$-greedy. Here, the parameters $\delta$ and $\gamma$ do the job (see \cite{hm2001}). We set these to $\delta = 0.1$ and $\gamma = 0.2$ for all experiments.\p

Finally, for WOLF-PHC, we need to specify two additional learning rates $\delta_w$ and $\delta_l$ with $\delta_l > \delta_w$ (see \cite{bv2002}). The algorithm uses the learning rate $\delta_w$ if it believes itself to be ``winning'', and it uses the rate $\delta_l$ if it believes itself to be ``losing''. We use a setting of $\delta_w = \frac{1}{1000 + t}$ and $\delta_l = 2 \delta_w$, where $t$ denotes the time (or current repetition) of the game.\p

We tested these settings in a series of experiments and found them to be working well. Nonetheless, the settings are likely to be sub-optimal. This, however, is irrelevant for our purposes since we are not considering the rate at which the algorithms learn about the action values or explore the environment. This can be neglected insofar as that an optimised parameter setting, when considered in the long-run, will not lead to fundamental improvements (such as an enhanced capability to learn Nash equilibria). Instead, we chose to use identical or similar parameter settings for all algorithms in order to simplify the analysis of the results.\p

	% Experimental results
	\section{Experimental results} \label{sec:exp-results}

\noindent This section presents and analyses the results of our experiments. It is important to note that the performance of an algorithm may depend on the player position it takes on. To account for this, we repeated each play once for every permutation of the agent order. We call this process a \emph{sweep}.\p

In the following, whenever we refer to statistical significance, this is based on a paired t-test with a significance level of 5\%. We use the notation ``Alg1 / Alg2'' if the performances of the algorithms Alg1 and Alg2 are statistically equivalent (i.e. the difference is statistically insignificant). All reported results are averaged over all plays, games, and teams.

		% No-conflict games
		\subsection{No-conflict games} \label{nocnf-games}

\noindent The first part of our experiments evaluated the algorithms in the set of all structurally distinct strictly ordinal $2 \times 2$ no-conflict games (21 games in total). We evaluated all combinations of the algorithms in each of the games. In total, we evaluated each pair of algorithms in 25 sweeps (50 plays), where each play consisted of 100,000 repetitions.\p

Table~\ref{tab:res-noconflict} shows the performance metrics for every algorithm. The maximum payoff any player can achieve in any game is 4, and the maximum social welfare and fairness, respectively, are 8 and 16. All algorithms performed quite well. We note that NashQ, both in homogeneous and heterogeneous teams, performed extremely well. A closer look at the strategy trajectories reveals that NashQ persuades the other agent to play a NE strategy by playing a NE strategy itself, regardless of whether it could achieve a higher payoff by using another strategy. However, NashQ requires more information than any of the other algorithms.\p

Next, we note that those algorithms that model their opponents (i.e. JAL and CJAL) perform generally better than those that do not (i.e. WOLF-PHC and RegMat, leaving NashQ aside). Both JAL and CJAL have better results than WOLF-PHC and RegMat throughout all constellations of agents. Note also that JAL and CJAL have almost identical performances (all significance tests were negative). The results indicate that opponent modelling techniques may be better suited for ad hoc team scenarios because they learn the strategies of the other agents irrespective of the algorithms on which they are based. This is in line with Barrett et al. \cite{bsk2011} and Wu et al. \cite{wzc2011}, whose algorithms are also based on opponent modelling techniques.\p

		% Conflict games
		\subsection{Conflict games} \label{sec:cnf-games}

\noindent The second part of our experiments evaluated the algorithms in the set of all structurally distinct strictly ordinal $2 \times 2$ conflict games (57 games in total). As before, we evaluated each combination of the algorithms in each of the games, using 25 sweeps and 100,000 repetitions per play. The results are shown in Table~\ref{tab:res-conflict}. The maximum payoff any player can achieve in any of these games is 4. However, the maximum welfare and fairness vary among the games and can be as high as 7 and 12, respectively.\p

First, we note that NashQ achieved the highest convergence rate, followed by CJAL / WOLF-PHC, then JAL, and then RegMat. The high convergence rate of NashQ is explained by the fact that it plays a NE strategy in each state, regardless of whether another strategy would provide higher payoffs. Therefore, as soon as NashQ has learned the payoff structure of the game, it will always play the same strategy. The low convergence rate of RegMat can be explained by the fact that it constantly tries to maintain (or restore) the Hannan consistency, forcing it to frequently change its strategy.\p

The final expected payoffs and the average welfare and fairness of all algorithms are very similar. Indeed, all of these are statistically equivalent. This is an interesting result since it does not correspond to the solution rates of the algorithms. Here, one can see that WOLF-PHC / JAL / RegMat have the highest NE rates, followed by NashQ and CJAL. On the other hand, CJAL has the highest PO rate, followed by all other algorithms with statistically equivalent PO rates. The same applies to the WO and FO rates, where CJAL again achieved the highest rate, while the other algorithms achieved equivalent rates. Furthermore, note that NashQ has the second lowest NE rate although it plays a NE strategy in every state. Other than in the previous section, playing a NE strategy in every state does not necessarily seem to persuade the other agent to play a NE strategy as well.\p

The results seem to indicate that CJAL may be better suited for ad hoc team problems than the other algorithms. It achieved the highest PO, WO, and FO rates, and these are significantly higher than those of the other algorithms. Moreover, it has the highest average welfare and fairness (yet these are statistically equivalent to the other algorithms). Note also that, apart from NashQ, it achieved the highest convergence rate. A high convergence rate is useful since it allows the other agents to adapt to a stationary opponent (that is, once it has converged). This is especially useful in ad hoc team scenarios because the agents may not always be able to resort to coordination strategies. However, note that CJAL has the lowest NE rate of all algorithms. Thus, whether CJAL is better suited for ad hoc team problems will ultimately depend on the solution concept that is considered most appropriate for this domain.\p

		% Random games
		\subsection{Random games} \label{sec:rnd-games}

\noindent In the third part of our experiments we investigated how well the algorithms scale to ad hoc teams with more than two agents. We used a modified version of the evaluation procedure proposed by Stone et al. (see Section~\ref{sec:games}). Specifically, we tested each of the algorithms in 500 randomly generated strictly ordinal $2 \times 2 \times 2$ games.\p

Table~\ref{tab:res-random} shows the performance metrics for every algorithm. The maximum payoff any player can achieve in any game is 8, while the maximum welfare and fairness may vary among the games. If the game happens to be a no-conflict game, then these values amount to 24 and 512, respectively. However, if the game is a conflict game, then the maximum welfare and fairness may assume any value as high as 23 and 448, respectively. Of all games generated in our experiments, 2\% were no-conflict games and 98\% were conflict games.\p

First, we note that NashQ has the highest convergence rate. This is because once the algorithm has learned the game structure, it will always play the same NE strategy. The second and third highest convergence rates were achieved by CJAL / WOLF-PHC and JAL / RegMat, respectively. It is interesting to see that CJAL / WOLF-PHC and JAL / RegMat have similar convergence rates, since, in both groups respectively, the first algorithm plays pure strategies while the second algorithm plays mixed strategies. One would expect the former to have a significantly lower convergence rate than the latter because those algorithms that play pure strategies need to change their strategies periodically in order to approximate mixed strategies.\p

The final expected payoffs (FEPs) of all algorithms are statistically equivalent. Note that NashQ achieved the third highest FEP. This is interesting because NashQ plays a NE strategy regardless of whether or not another strategy provides higher payoffs. Moreover, it is remarkable that JAL, CJAL, WOLF-PHC, and RegMat have equivalent FEPs despite the fact that the former two play pure strategies while the latter two play mixed strategies. This could indicate that the impact on the average payoffs due to the constellation of agents in a team may not be as strong as one might otherwise expect.\p

The average welfare and fairness confirm our observations of Section~\ref{sec:cnf-games}. CJAL achieved a higher welfare and fairness than any other algorithm. The welfare and fairness of the other algorithms are statistically equivalent. This corresponds to the WO and FO rates of the algorithms, where CJAL achieved significantly higher rates than all other algorithms. In fact, in almost half of all plays, CJAL managed to arrive at a welfare-optimal solution, of which a majority was fairness-optimal as well. As noted earlier, this may be a valuable property for ad hoc team problems. An ad hoc agent that is able to collaborate with an unknown group of agents such that the overall performance of the entire group is optimised (in terms of welfare and fairness) may be better than an agent that attempts to optimise its own payoff only. However, we note again that this essentially depends on the priorities of both the ad hoc agent and the entire group.

\vspace{0.05cm}

Finally, consider the different solution rates. The highest NE rate was achieved by RegMat, followed by WOLF-PHC, JAL / NashQ, and then CJAL. Note that the NE rate of CJAL is relatively low when compared to the other agents. However, this is opposed by the PO rates. Here, CJAL / NashQ achieved the highest rate, followed by JAL / RegMat and WOLF-PHC. It is interesting that CJAL and NashQ achieved equivalent PO rates, despite the fact that CJAL was specifically designed to learn PO solutions \cite{bv2002}, whereas NashQ was specifically designed to learn NE solutions \cite{hw2003}. Note also that the PO rate of WOLF-PHC is quite low, especially in comparison to its relatively high NE rate.\p

		% Overall results
		\subsection{Overall results}

\begin{figure}[h]
	\centering
	\includegraphics[width=0.47\textwidth]{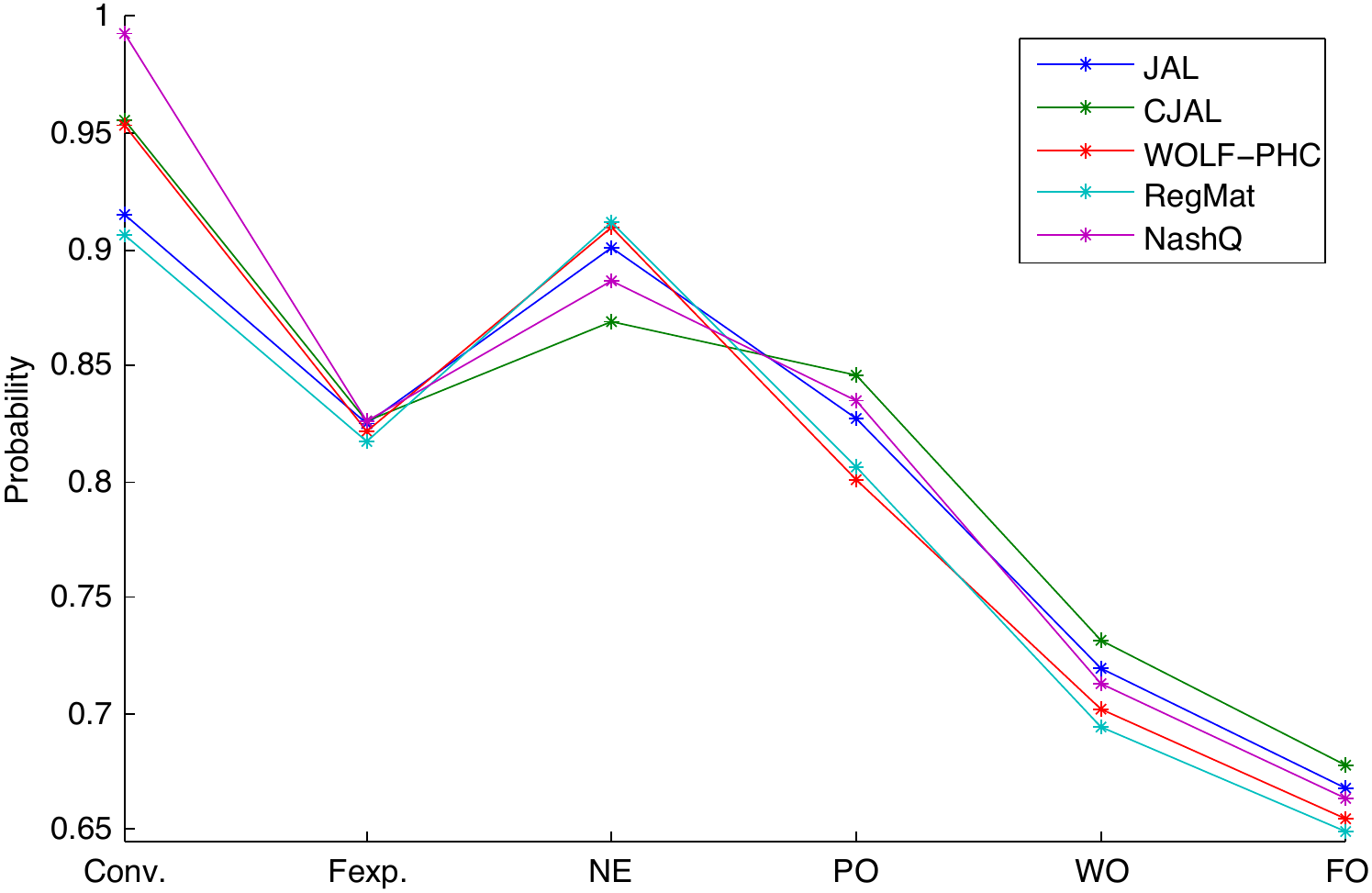}
	\caption{Overall results.}
	\label{fig:overall-results}
\end{figure}

\noindent Figure~\ref{fig:overall-results} shows the results averaged over all three parts. Note that we cannot take the average of the payoffs since we consider ordinal games. However, for the final expected payoffs, we first normalised the values by dividing through the respective maximum, after which we took the average of all results. Thus, the final expected payoffs in Figure~\ref{fig:overall-results} are to be read as percentages where 0\% means that the algorithm always achieved its least preferred outcome, and 100\% means that it always achieved its most preferred outcome.\p

The results show that there is, in fact, no algorithm which achieved an overall better performance in the experiments, that is, \emph{no algorithm is generally better}. The following summary underlines this:\p

\begin{itemize}
	\item JAL has the second lowest convergence rate with about 91.47\%. This comes from the fact that it changes its strategy periodically in order to approximate a mixed strategy. Furthermore, it has the third highest payoff rate with 82.49\%. This rate is statistically equivalent to the highest rates. The NE rate of JAL is the third highest with about 90\%. Finally, it has the third highest PO rate (82.74\%), the second highest WO rate (72\%), and the second highest FO rate (66.81\%).\p

\vspace{0.23cm}

	\item CJAL has the second highest convergence rate with about 95.59\%. This is since it attempts to learn Pareto-optimal solutions, which, in our case, are often pure profiles. Moreover, of all algorithms used in our experiments, it has the highest PO rate (84.56\%), WO rate (73.14\%), and FO rate (67.76\%), which consequently leads to the highest payoff rate with about 82.57\%. However, this is opposed by the fact that it achieved the lowest NE rate with 86.9\%.\p

	\item WOLF-PHC has the third highest convergence rate with about 95.35\%. This is statistically equivalent to CJAL. Furthermore, it has the second lowest payoff rate with 82.12\%. On the other hand, with about 90.96\% it has the second highest NE rate. This rate is statistically equivalent to the highest NE rate. Finally, it has the second lowest PO rate (80.12\%), WO rate (70.17\%), and FO rate (65.53\%).\p

	\item RegMat achieved the lowest convergence rate with 90.61\%. As pointed out earlier, it frequently changes its strategy in order to maintain the Hannan consistency. Moreover, with about 81.72\%, it also has the lowest payoff rate. This is a consequence of its frequent changes. Interestingly, these efforts lead to the highest NE rate of all algorithms with about 91.14\%. However, they also lead to the lowest PO rate (80.66\%), WO rate (69.46\%), and FO rate (64.95\%).\p

	\item NashQ, with about 99.26\%, managed to converge in most of the games. As was explained earlier, this is since it will always play the same strategy after it learned the payoff structure of the game. It is worth noting that it has the second highest payoff rate with about 82.56\%, which is almost identical to the highest rate. This is interesting since NashQ chooses its strategies irrespective of the strategies of the other agents. On the other hand, it is surprising that it has the second lowest NE rate (88.59\%), despite the fact that it plays a NE strategy in each state. It is equally surprising that it achieved the second highest PO rate (83.49\%) and the third highest WO rate (71.35\%) and FO rate (66.39\%), despite the fact that it was not optimised for these solution types.\p
\end{itemize}

We conclude that the assessment of an algorithm ultimately depends on the solution concept that is considered most appropriate for the problem at hand. In other words, its performance depends on the priorities of the entire team. For example, in the predator domain investigated by Barrett et al. \cite{bsk2011}, it would be most desirable to arrive at a welfare-optimal solution if we define the welfare of the predator group to be the inverse of the average steps needed to capture the prey. Moreover, if we think of the agents as real robots (e.g. as in \cite{sk2010}), then we might want to arrive at a fairness-optimal solution in order to ensure that all robots have identical or similar energy consumptions. On the other hand, in a multiagent marketing application in which the other agents cannot be trusted (as they may want to deceive us in order to increase their payoffs), we would want to arrive at a Nash equilibrium (or minimax profile in 2-player zero-sum games) such that we can guarantee a minimum average payoff.\p

Although this conclusion may seem unsurprising at first glance, the implications should be of interest to researchers developing multiagent learning algorithms. The typical focus in this area has been on the concept of Nash equilibria (or, equivalently, minimax profiles in 2-player zero-sum games). For a selection, see \cite{cb1998,bv2002,hw1998,hw2000,hw2003,uv1997,b1951,lh1964,l1994,ls1996,l2001,cs2003}. Some authors focus on the concept of Pareto optimality as an alternative \cite{sam2003,kl2005,bs2007}, others focus on correlated equilibria \cite{hm2000,hm2001,gh2003}, and some do not make any specific commitments regarding the nature of the solution \cite{sk2010,skr2010,bsk2011,wzc2011}. However, as can be seen from our results, it is important to consider a wider spectrum of solution concepts in order to fully assess the performance of an algorithm. Indeed, this bears an interesting resemblance to the No Free Lunch theorems of Wolpert and Macready \cite{wm1995,wm1997}. Therein, roughly speaking, it is argued that the performance of any two algorithms is identical when averaged over all possible problems. That is, whenever an algorithm is superior to another algorithm on a certain set of problems, this is paid for by inferiority on a different set of problems. Our results show a tradeoff relation of this kind. For instance, CJAL has the highest PO rate and the lowest NE rate whereas RegMat has the lowest PO rate and the highest NE rate. Other algorithms range somewhere in the middle, without best or worst performances. This seems to indicate that superiority in one solution type is compensated for by a converse relation somewhere else.\p

	% Related work
	\section{Related work} \label{sec:rel-work}

\noindent Harsanyi pioneered the study of incomplete information games. In his 1967 paper \cite{h1967}, he describes the \emph{Bayesian game}, a game in which players have beliefs about missing information. He develops the concept of the Bayesian Nash equilibrium \cite{h1968a} in which each player plays a best response against the other players, based on the personal beliefs of the player. Jordan \cite{j1991} showed that, for any repeated game, if the players play a Bayesian Nash equilibrium in each repetition, and if the personal beliefs of the players satisfy certain conditions, then this will converge to a true Nash equilibrium.\p

The problem of incomplete information in multiagent learning, in the form of the ad hoc team problem, was addressed by Stone et al. \cite{skkr2010}. They propose a procedure to evaluate two ad hoc agents for a given set of potential team members and tasks. We used a modified version of this procedure for our own experiments (see Section~\ref{sec:games}).\p

In earlier work, Stone and Kraus \cite{sk2010} define optimal strategies for an ad hoc agent collaborating with a fixed-behaviour teammate in an environment modelled as a $k$-armed bandit. Stone et al. \cite{skr2010} present an algorithm that would lead a fixed greedy agent towards an optimal joint action in a simple repeated game in which both agents have identical payoff functions.\p

More recently, Genter et al. \cite{gas2011} introduced a framework for a \emph{role-based} approach to the ad hoc team problem. Using a set of predefined roles (i.e. behaviours), the ad hoc agent tries to assume a role such that the marginal utility of the team is maximised. The agent was shown to be effective in several instances of the Pac-Man domain. However, we note that the framework is based on a number of key assumptions. It assumes that all teammates follow one of a finite set of a priori predefined roles, that the ad hoc agent knows what roles its teammates follow, and that the ad hoc agent knows the internal payoff distributions of its teammates.\p

These assumptions are relaxed in a recent empirical study by Barrett et al. \cite{bsk2011}. They used an ad hoc agent that tries to identify its teammates by observing their behaviour and comparing it with a database of known behaviours. In addition, it learns a new model for the observed behaviour using a tree classifier. The agent combines both the database and the learned model in a Bayesian fashion to anticipate the behaviour of its teammates. Experiments showed that the ad hoc agent performed quite well, and in general better than those agents that just mimic their teammates.\p

Wu et al. \cite{wzc2011} proposed an interesting algorithm called \textit{Online Planning for Ad Hoc Agent Teams} (OPAT). For each encountered state, the algorithm estimates the values of all joint actions using Monte-Carlo Tree Search. These values are used to generate a stage game (i.e. a repeated game with one repetition), based on which the algorithm decides which action to take. The decision process considers the past $m$ plays of the current stage game to approximate the strategies of the other agents. OPAT was shown to be effective in a series of multiagent domains.\p

	% Conclusion
	\section{Conclusion} \label{sec:conclusion}

\noindent In this work, we compared the performance of five multiagent learning algorithms in a set of ad hoc team problems. The algorithms were evaluated in a comprehensive range of repeated games, and the teams consisted of agents which were themselves learning. Our intention was to characterise the performance of salient types of multiagent learning algorithms in ad hoc team problems. Our experiments show that there is no clear favourite among the algorithms. In particular, we conclude that the performance of an algorithm ultimately depends on the solution concept that is considered most appropriate for the problem at hand.\p

The experiments in this paper were based on 2-player and 3-player matrix games, in order to make comparative statements in a well defined and comprehensive set of game types. It would be of interest to extend this analysis to the case of multi-player games (i.e. $n$-player games with $n > 3$), and to games with multiple states (i.e. stochastic games \cite{s1953}). We expect that such extensions would bring many known difficulties regarding interactive decision making \cite{y2004} to the fore and perhaps differentiate the multiagent learning algorithms further. We anticipate that going down this path may also clarify when one may need to draw on more elaborate models than stochastic games, e.g., as in \cite{s1980}.\p

	% Bibliography
	\newpage
	\small
	\bibliographystyle{abbrv}
	\bibliography{aamas12}

\end{document}